\magnification=1200
\pageno=1

\centerline {\bf THE SPINNING MEMBRANE  AND }
\centerline {\bf SKYRMIONS REVISITED}
\bigskip
\centerline { Carlos Castro}
\centerline { Center for Theoretical Sudies of Physical Systems}
\centerline { Clark Atlanta University, Atlanta, GA.30314}
\smallskip
\centerline { June 1997}
\bigskip
\bigskip

\centerline { \bf ABSTRACT}

A local world volume Q-supersymmetric Weyl invariant Lagrangian for the membrane is presented. 
An analysis is provided  which solves the problems raised by some 
authors in the past concerning the algebraic elimination of the auxiliary fields belonging to the coupling function supermultiplet. 
The starting bosonic action is the one given by 
Dolan and Tchrakian
with vanishing cosmological constant and with quadratic, quartic derivative terms. 
Our Lagrangian differs from 
the one of Lindstrom and Rocek in the fact that is polynomial in the fields facilitating the quantization 
process. It is argued, rigorously, that if one wishes to construct polynomial actions without curvature 
terms and where supersymmetry is linearly realized, after the elimination of auxiliary fields, one must 
relinquish S supersymmetry and concentrate solely on the Q-supersymmetry associated with the 
superconformal 
algebra in three dimensions. 
The role that this spinning membrane action  may have in the theory of D-branes, 
Skyrmions and BPS monopoles is also pointed out.

\bigskip

\centerline{\bf I. {INTRODUCTION}}
\bigskip

      In the past years there has been considerable
progress in the theory of extended objects, in particular two dimensional extended
objects ; i.e.  membranes. However, a satisfactory spinning
membrane Lagrangian has not been constructed yet, as far
as we know. Satisfactory in the sense that a suitable
action must be one which is polynomial in the fields,
without ( curvature) $R$  terms which interfere with the algebraic
elimination of the three-metric, and also where
supersymmetry is linearly realized in the space of
physical fields. It had been argued [1,2] that it was  allegedly 
impossible to
supersymmetrize Dirac-Nambu-Goto type of actions
(DNG)
-those proportional to the world-sheet and
world-volume spanned by the string (membrane) in their
motion through an embedding space-time. The efforts to
supersymmetrize this action have generally been based upon
the use of the standard, classically-equivalent, bosonic
action which included a cosmological constant. The
supposed obstruction is related to the fact that in order
to supersymmetrize this constant one had to include an 
Einstein-Hilbert term spoiling the process of the algebraic elimination of the three-metric altogether.

       Bergshoeff  et al [ 3] went evenfurther and
presented us with the "no-go" theorem for the spinning
membrane. Their finding was based in the study of a
family of actions, in addition to the one comprised of
the cosmological constant, which were equivalent, at the
classical level, to the DNG action. However, this "no-go"
theorem was flawed because these authors relied on the
tensor calculus for Poincare D=3 N=1 SG developed by
Uematsu [ 4  ]. Unfortunately, the above tensor calculus despite being correct 
does $not$ even yield a linearly realized supersymmetry for
the kinetic matter multiplet to start with!. A constraint,
 $\bar{\chi}\chi$ =0  ,  appears after the elimination
of the $S$ auxiliary field, where $\chi$
is the $3D$ Majorna spinor. The spinning membrane requires supersymmetry on the world  volume 
whereas the supermembrane requires  target space-time supersymmetry. 
Lindstrom and Rocek [5]  were the first ones to
construct a Weyl invariant spinning membrane action.
However, such action was highly non-polynomial
complicating the quantization process evenfurther than the
one for the supermembrane and the membrane coordinates had a non-canonical dimension from the world volume 
point of view. 

     The suitable action to supersymmetrize is the one of
Dolan and Tchrakian [6] ( DT) without a cosmological constant
and with quadratic and quartic-derivative terms. 
Such membrane action is basically a Skyrmion  action with quartic and quadratic derivative terms. 
A class
of conformally-invariant $\sigma$- model actions was
shown to be equivalent, at the classical level, to the
DNG action for a p+1 extended object ( p+1=even) embedded
in a target spacetime of dimension $ d\geq {p+1}$. When
p+1=odd, our case, an equivalent action was also
constructed, however, conformal invariance was lost in
this case. The crux  of the work presented here  lies in the necessity to embed the Dolan-Tchrakian action into 
an explicitly Weyl invariant one through the introduction of extra fields. These are the gauge field of 
dilations , $b_\mu$, not to be confused with the $U(1)$ world-volume gauge field appearing in $D$-branes, and 
the scalar coupling ( a conformal compensator field ) , $A_0$ of dimension $(length)^3$, that must appear in front of the quartic 
derivative terms of the DT action. The latter terms must appear with a suitable coupling constant in order to 
render the action dimensionless. As a result of the embedding into a Weyl invarant action the coupling constant 
turns into a scalar of Weyl weight equal to $-3$. A similar procedure occurs in the Brans-Dicke formulation 
of gravity.

Having Weyl-covariantized the DT action, the natural question to ask is how do we eliminate these
new fields, $b_\mu, A_0$ in order to recover the original action ? This can be achieved simultaneously
if one imposes the constraint, $D^{Weyl}_\mu A_0 =0$ which implies that the gauge field $b_\mu $ is pure 
gauge : $b_u \sim \partial_\mu lnA_0$ so that when one fixes the dilational gauge invariance by choosing 
: $A_0=g$, ($g $ constant)
 the gauge field becomes $b_\mu =0$. There  might be global topological obstructions to gauge to zero the $b_\mu$ everywhere  
that are not discussed here [7] but that might be very relevant in the nonperturbative behaviour of the theory. 
The constraint $D^{Weyl}_\mu A_0 =0$ can be $derived$ from first principles; i.e. from an action as it is shown 
in the appendix. It follows that if the equations of motion of the Weyl covariantized  DT action are indeed the
Weyl covariant extension of the DT equations of motion then one must have  $D^{Weyl}_\mu A_0 =0$ which is 
tantamount of saying that the coupling scalar $A_0$ is just the analog of a constant : 
the Weyl covariant extension of the original 
( coupling)  constant. Therefore, on-shell dilatational gauge invariance of the Weyl covariantized DT action 
(WCDT) allows to recover the original DT action upon choosing the gauge $A_0=g$ and implementing  the embedding condition 
 $D^{Weyl}_\mu A_0 =0$. The superconformal extension requires to use  the superconformal covariant constraint 
$D^C_\mu A_0 =0$ where the superconformal 
covariant derivative is given in the text. 
The off-shell extension requires 
further study especially in so far  the quantization process. 

Once the embedding program  into the WCDT action has been performed one supersymmetrizes the WCDT action by 
incorporating $A_0$ into the superconformal coupling-function multiplet $(A_0, \chi_0, F_0)$, whereas the $b_\mu$
becomes part of the gravitational conformal supermultiplet involving $(e^m_\mu, \psi_\mu, b_\mu)$ and the physical fields 
of the membrane form part of the world-volume superconformal multiplet $(A^i, \chi^i, F^i)$. 
The $A^i$ fields are identified with the membrane target space time coordinates.

If one wishes to eliminate any curvature scalar terms in the final action  one must take 
suitable combinations of these three superconformal multiplets  and, in doing so, one is going to break  
$explicitly$ the $S$ supersymmetry of the $3D$ superconformal algebra as well as the conformal boost symmetry, the $K$ symmetry,  
which signals the
$presence $ of the $b_\mu$ field in the final action : it does $not$ decouple as it does in ordinary 
$3D,4D$ superconformal gravitational actions where the $full$ superconformal invariance is maintained that 
allows to fix the conformal boost $K$-symmetry by choosing the gauge condition  $b_\mu =0$.

 The final action is Lorentz, dilational, $Q$ supersymmetric and translational invariant but is $not$ 
invariant under $S$ supersymmetry nor conformal boosts, $K$ . There is nothing 
$wrong$ with this fact because the subalgebra  of the full $3D$ superconformal algebra comprised of the Lorentz
generator, dilations, $Q$ supersymmetry and translations, $P_\mu$, does $close$!. In conventional Poincare 
supergravity one has invariance only under a particular linear combination of $Q$ and $S$ supersymmetry, the 
so called  $Q+S$ sum rule and the original $K$ symmetry is used to  enforce the decoupling of the $b_\mu$ from the 
action  
by fixing the conformal boost symmetry  $b_\mu =0$. 
Here we have a different picture, we have full $Q$ supersymmetry instead of a particular combination of $Q$ and $S$ 
and there is  $no$ explicit conformal boost invariance to start with. 

In the past there has been a lot of debate concerning the elimination of the auxiliary fields, $F^i,F_0$ [8]. 
We have shown that there are $no$ constraints ( which will spoil the linear realization of supersymmetry in the Hilbert space of physical fields)  among the physical membrane fields after the elimination of the auxiliary fields $F^i$. 
Earlier in [8] we discussed what happens upon the elimination of $F_0$. Since supersymmetry rotates 
field equations into field equations among given  members of a supermultiplet, one $must$ and $should$ 
eliminate as well the remaining members of the coupling supermultiplet , $A_0,\chi_0$. When this is done the $full$ 
quartic supersymmetric terms are constrained to zero. This does $not$ imply that the quartic terms in the bosonic sector are zero; it is the 
whole sum of the bosonic quartic terms and their supersymmetrization   which is constrained to zero. 
This constraint is $unnatural$ because the coupling multiplet must $not$ be treated as a Lagrange multiplier. Secondly,  upon setting the fermions to zero the 
auxiliary field $F_0$ decouples from the action any way [8] and , thirdly, if one varies the couplings while 
maintaining the embedding condition, $D^c_\mu A_0 =0$,   one $cannot$ longer vary  simultaneously the 
members of the 
gravitational super multiplet as one should in order to retrieve the Dirac-Nambu-Goto actions. Because if 
this is done too many 
restrictions will arise among the three supermultiplets.

The correct procedure ( instead of the one in [8]) , due to the fact that the $b_\mu$ field 
does $not$ decouple from the action and also because it is a member of the supermultiplet 
containg the graviton and gravitino, is to vary those members of the gravitational supermultiplet and $not$ those  
of the coupling function supermultiplet. 
The superconformal covariant embedding condition, $D^c_\mu A_0 =0$, plus its series of supersymmetric transformations  
( ``rotations'') , will furnish the desired relations between the 
fields $(A_0,\chi_0,F_0)$ 
and the $3D$ background gravitational supermultiplet $(e^m_\mu, \psi_\mu, b_\mu)$ which,  in turn , can be 
determined in terms of the physical membrane fields,  $(A^i, \chi^i, F^i)$  resulting from  the $algebraic$ 
elimination of the gravitational supermultiplet ( non-propagating field equations)  upon the variation of the action.   

Therefore, concluding,  the background gravitational supermultiplet and the coupling supermultiplet are fully 
determined in 
terms of the membrane physical fields after using the embedding condition  $D^c_\mu A_0 =0$, 
and eliminating the members of the gravitational superconformal multiplet via their nonpropagating equations of motion.  
Thus there is $no$ need to vary the fields $F_0, A_0,\chi_0$ in order to eliminate them algebraically as we did in [8]. It is sufficient to vary the gravitational supermultiplet supplemented by the superconformal covariant embedding condition  
 $D^c_\mu A_0 =0$. This is one of the main new points we wished to add to the spinning membrane literature.

The outline of this work goes as follows. In the first part of section II we present the work of Dolan and
Tchrakian and discuss the problems
associated with the $3D$ Poincare Supergravity tensor calculus : 
a constraint arises among the physical fields 
upon elimination of the auxiliary fields. This is why 
the "no-go" theorem was inappropriate for $3D$ Poincare Supergravity. 
In the last part of {\bf II} the connection between the supersymmetric version of the generalized Skyrme's   model of baryons  
and  the   
spinning membrane propagating in a curved backgrounds is studied and the role that the 
spinning membrane may have in the physics of D-branes and BPS monopoles is discussed. It is suggested that $D0$-branes may play an important role in the spectrum of the generalized Skyrmion action .    

In III, we provide   the detailed  arguments showing that in
order to satisfy all of the stringent requirements
discussed earlier in order to have a satisfactory spinning membrane we must relinquish $S$ 
supersymmetry and
concentrate solely on the $Q$ supersymmetry associated with
the superconformal algebra in three dimensions. $"Q+S"$ 
supersymmetry can only be implemented in the class of
non-polynomial actions (if we insist in meeting all of our
requirements) as it was shown in  [5] . The fully
$Q$ invariant action is furnished providing  a $Q$-spinning  and Weyl invariant  membrane .
Finally, upon the elimination of the 
auxiliary fields $F^i$ no constraints arise among the membrane fields.

To finalize, in the first part of the appendix the we show that the $b_\mu$ field does $not$ decouple from the action and the elimination 
from its algebraic field equations ( instead of the $F_0$ field )  solves 
the problems  raised by some authors in the past.  
In the final part of the appendix the embedding condition  $D^c_\mu A_0 =0$ is derived from first principles. 
The latter condition is essential in order to retrieve the Dolan-Tchrakian action after the elimination of the 
auxiliary $F^i$ fields, setting the fermions to zero and fixing, finally,  the gauge $A_0 =g$.

     Our coventions are: Greek indices stand for
three-dimensional ones; Latin indices for spacetime ones : $i,j =0,1,2....D$. The signature of the $3D$ volume is 
$(-,+,+)$. 

\bigskip
\centerline{\bf II}
\smallskip

\centerline {\bf 2.1 The Dolan-Tchrakian Action}

The Lagrangian for the bosonic $p$-brane ( extendon) with vanishing cosmological constant constructed by Dolan and
Tchrakian in the case that $p=odd; p+1=2n$ is :

$$L_{2n} =\sqrt {-g} g^{\mu_1 \nu_1}.... g^{\mu_n \nu_n} \partial_{[\mu_1} X^{i_1}.........\partial_{\mu_n]} X^{i_n}
\partial_{[\nu_1} X^{j_1}.........\partial_{\nu_n]} X^{j_n} \eta_{i_1 j_1}.... \eta_{i_n j_n}. \eqno (2-1)$$
$\eta_{ij}$ is the spacetime metric and $g^{\mu\nu}$ is the world volume metric of the $2n$ hypersurface spanned by the motion
of the $p$-brane. Antisymmetrization of indices is also required. Upon elimination of the $2n\times 2n$ matrix :

$$ A^\mu_\nu =g^{\mu \rho} \partial_{\rho} X^{i}(\sigma)\partial_{\nu} X^{j}(\sigma)\eta_{ij}. \eqno (2-2) $$
from the $n^{th}$ order polynomial in $A^\mu_\nu$:

$$ A^n -b_{2n-1}A^{n-1} +b_{2n-2}A^{n-2}..................(-1)^{n+1} b_{n+1}A+{1\over 2} (-1)^n b_n I_{2n\times 2n}=0. \eqno (2-3) $$
where  the scalar coefficients in the matrix-polynomial equation are obtained from the first $n+1$ coefficients in 
the expansion of :

$$det(A^\mu_\nu -\lambda I_{2n\times 2n}) =\lambda^{2n} -b_{2n-1} \lambda^{2n-1}+b_{2n-2} 
\lambda^{2n-2}..........-b_1\lambda +det(A^\mu_\nu). \eqno (2-4)$$

A substitution of the matrix solution to the matrix-polynomial equation (2-3) with coefficients given by (2-4) into the action (2-1) yields :

$$L_{2n} =\sqrt {-det( \partial_{\mu} X^{i}(\sigma)\partial_{\nu} X^{j}(\sigma)\eta_{ij})}~[{(n!)^2 b_n \over \sqrt A}]. \eqno (2-5) $$
The crucial observation made by [6] is that the last factor :

$$[{(n!)^2 b_n  \over \sqrt {A}}].~~~A=det~A^\mu_\nu. \eqno (2-6)$$
takes discrete values for all values of $n$. Therefore , the equivalence to the Dirac-Nambu-Goto action has been established. 
Notice that for every $n$, $L_{2n}$ is conformal invariant and it is only quadratic in time derivatives due to the antisymmetry of the indices. Therefore 
attempts to quantization might not be hopeless.  

When $p=even$ , $p+1$ odd,  a Lagrangian with zero cosmological constant can also be constructed, however, conformal invariance is $lost$.
In the membrane's case one has :

$$L=L_4 +aL_2.~~~a>0. ~~~L_2 =\sqrt {-g} g^{\mu\nu}\partial_{\mu} X^{i}(\sigma)\partial_{\nu} X^{j}(\sigma)\eta_{ij}. $$
$$L_4 =\sqrt {-g} g^{\mu\nu}  g^{\rho \tau}\partial_{[\mu} X^{i}(\sigma)\partial_{\rho ]} X^{k}(\sigma)\partial_{[\nu} X^{j}(\sigma)\partial_{\tau ]} X^{l}(\sigma)
\eta_{ij} \eta_{kl}             . \eqno (2-7) $$

Upon elimination of the world volume metric one gets : 

$$12 \sqrt {a} \sqrt {-det G_{\mu\nu}}~or~  -4\sqrt {a} \sqrt {-det G_{\mu\nu}}. \eqno (2-8)$$
with 

$$G_{\mu\nu}= \partial_{\mu} X^{i}(\sigma)\partial_{\nu} X^{j}(\sigma)\eta_{ij}. \eqno (2-9) $$
Notice that $a >0$ in (2-7) so both $L_2$ and $L_4$ have the same relative sign.  

\smallskip

\centerline {\bf 2.2 A nonlinearly realized Poincare-supersymmetric Membrane}
\smallskip

The Poincare-supersymmetric kinetic terms ( modulo total derivatives) for the $3D$ Poincare supergravity 
was given by Uematsu [4]. The fields are $\Sigma_P=(A,\chi, F')$ where one must not confuse the 
auxiliary field $F'$ with the one of the superconformal multiplet : $F=F'+{1\over 4} AS$. 
An invariant action can be constructed using the tensor calculus [4] from the supermultiplet $\Sigma_P$ and its 
kinetic scalar multiplet $T_P(\Sigma_P)$ as follows  :

$$L={1\over 2} [\Sigma_P \otimes T(\Sigma_P)]_{inv} -{1\over 4} [T(\Sigma_P \otimes \Sigma_P)]_{inv}=$$

$$e[-{1\over 2} g^{\mu\nu} \partial_\mu A \partial_\nu A -{1\over 2} {\bar \chi} \gamma^\mu D_\mu \chi +
{1\over 2} F'^2 +{1\over 2}  {\bar \psi}_\nu \gamma^\mu \gamma^\nu \chi \partial _\mu A +$$
$$
{1\over 16} {\bar \chi}\chi {\bar \psi}_\nu \gamma^\mu \gamma^\nu\psi_\mu +{1\over 8} S{\bar \chi}   \chi].
\eqno ( 2-10)$$
The Lagrangian is essentially identical to the Neveu-Ramond-Schwarz spinning string with the crucial difference that 
the ``effective mass'' term $S{\bar \chi} \chi$ is $not$ present in the string case ! 
Therefore upon eliminating the auxiliary field $S$ in the action (2-10) yields the unwanted constraint :  
$S{\bar \chi} \chi=0$ that spoils the linear realization of supersymmetry to start with ! In order to remedy 
this one could add the pure supergravity action with a corresponding $S^2$ term; however, this is precisely
what one wanted to avoid : the presence of $R$ terms in our action. 
Despite being able to write down a Poincare-supersymmetric extension of the DT action one still will be 
faced with the problem that supersymmetry will not be linearly realized upon elimination of the 
$S$ auxiliary fields.

There are ways to circumvent this 
problem. One way was achieved by Linstrom and Rocek who started with a non-polynomial  
Weyl invariant action (  if the membrane coordinates had a non-canonical Weyl weight of zero )  :

$$I\sim \int d^3\sigma~\sqrt {-g} [g^{\mu\nu}\partial_\mu X^i \partial _\nu X^j \eta_{ij}]^{3/2}.\eqno(2-11)$$
Since the auxiliary field $S$ is an alien concept in conformal supergravity it cannot appear  in the supersymmetrization process
where one uses conformal supergravity techniques to build invariant actions. The coordinates in 
(2-11) $X^\mu$ have the same dimension as their two-dimensional (string) counterparts. 
Since the action is non-polynomial the quantization process is hampered considerably. 
For this reason we must look for another 
option and supersymmetrize  the Dolan-Tchrakian action at the expense of introducing the $3D$ world volume 
gauge field of dilations $b_\mu$ and relinquishing $S$ supersymmetry and conformal boost invariance as well.

\bigskip
\centerline {\bf 2.3 D-branes, Skyrmions and BPS monopoles} 
\bigskip

The Dolan-Tchrakian action for the membrane is equivalent to the generalized Skyrme  action 
discussed among others by Manton [11]. We will follow Manton's work closely . 
A Skyrmion may be regarded as a topologically non trivial map from one Riemannian 
manifold to another minimizing a particular energy functional; i.e. classical static field configuration of minimal energy in a nonlinear scalar field theory, the pion field. The standard Skyrmion represents the 
baryon and has a conserved topological charge which precisely prevents a proton from decaying into pions. 
The charge is identified with the conserved baryon number or the degree of the map from 
$R^3 \rightarrow SU(2)$.  

Manton has emphasized that there  is no need for the target manifold to be a Lie Group. Take a map  
$\pi$ from $\Sigma_0 \rightarrow \Sigma_1$. Assume that both base and target space are three dimensional.
The three frame vectors $E^i_m, m=1,2,3 $ on $\Sigma_0$  are mapped by $\pi$ to the vectors 
$E^i_m \partial _i \pi^\alpha =
E^\alpha_m $ on $\Sigma_1$. 
The quadratic terms of the Skyrme model/ Dolan-Tchrakian action ( up to a minus sign) are just
the measure of how the sum of the squared-lenghts of frame vectors changes under the map $\pi$. 
The quartic terms corresponds of how the norm-squared of the area-elements constructed from the 
frame vectors change under the map. The equivalence is established once the corresponding indices are 
properly matched as :

$$ g^{mn} E^i_m \partial _i \pi^\alpha E^j_n \partial _j \pi^\beta g_{\alpha \beta} 
            \leftrightarrow g^{\mu\nu}\partial_{\mu} X^{i}(\sigma)\partial_{\nu} X^{j}(\sigma)\eta_{ij}. 
\eqno (2-12a)$$
similarly the norm of the area elements corresponds to the quartic terms :

$$|E^i_m \partial _i \pi^\alpha \wedge E^j_n \partial _j \pi^\beta|^2 
\leftrightarrow g^{\mu\nu} g^{\rho\tau}\partial_{[\mu} X^{i}(\sigma)\partial_{\rho]} X^{k}(\sigma)
\partial_{[\nu} X^{j}(\sigma)\partial_{\tau]} X^{l}(\sigma)\eta_{ij} \eta_{kl}.\eqno (2-12b)
$$
Eqs-(2-12) have a similar structure to the bosonic terms ( excluding the zero modes ) of the 
lightcone spherical supermembrane moving in a flat target spacetime background : a Yang-Mills theory of the 
area-preserving diffeomorphisms dimensionally reduced to one temporal dimension : a matrix model [2,12].  
The area-squared 
terms are just the same form as the $ [X^I,X^J]^2$ elements  appearing in the light-cone spherical bosonic membrane 
action whereas the length squared terms correspond to the 
ordinary kinetic energy terms. The main difference with eqs-(2-12a, 2-12b) is that $no$ gauge has been fixed, Lorentz covariance is fully explicit and manifest.

The total energy of the generalized Skyrmion is given by the Dolan-Tchrakian action up to an overall 
minus sign. 
For more details we refer to [11]. The advantage of the equivalence in eqs-(2-12) is that Lorentz invariance is manifest in contradistinction of the model of [12] that required the infinite momentum frame. Secondly, the model in [12] is 
background dependent whereas one must have a background independent formulation of nonperturbative string theory. 
Thirdly, one is not forced to take the $N=\infty$ limit in (2-12). Manton has discussed as well how to generalized Skyrme
model to $SU(N)$ for example. 
Duff et al [11] had also discussed in the past the behaviour of the actions like (2-7) and (2-11) which violated 
Derrick's theorem. Stable, static, nonsingular, finite energy kink  solutions exist in $3D$ once 
nonstandard actions like the Skyrmion action are built.

It would be interesting to incorporate  the eight transverse bosonic degrees of freedom of the 
$11D$ supermembrane 
as the dynamical variables of the  
target $SU(3)$ Lie group manifold in the Skyrme model with the purpose of attempting to relate QCD to membranes, if 
possible [8]. Other supermembrane target spacetime dimensions do $not$ fit into groups of the type $SU(N)$.    
If one wishes one can  supersymmetrize the generalized Skyrme theory exactly the same way we are going to 
do with the DT 
action in the next section. Hence, one has in one scoop a supersymmetrization of the Skyrme model via the 
DT action and the correspondence given by eqs-(2-12).

Now we turn to the BPS states. As remarked by Manton et al [11] , in recent work,  
there is mounting evidence that there is a close connection between $SU(2)$ BPS monopoles and Skyrmions. 
BPS monopoles are 
solutions of the Bogomolny equation that minimize 
classical energy solutions to the Yang-Mills-Higgs theory.  
Many low energy solutions to Skyrme's  equation look like monopoles with the baryon number 
identified with monopole number. The fields are not the same but the energy density configurations have 
equivalent symmetries 
and approximately the same spatial distribution . 

There is also a deep connection between BPS states and D-branes [9]. $D$-branes are essentially topological defects ( domain walls) where open-strings ends can move. Both in type II and heterotic string 
one can find BPS states in the perturbative string spectrum. 
It is believed that all perturbative string theories are diffferent faces of one underlying theory 
. This is known as string duality [10], where all these theories are believe to be different 
perturbative expansions of one underlying 
theory around different points in the moduli space of string vacua : 
In particular, type II superstring theories have supersymmetric $p$-brane solitonic solutions supported by 
Ramond-Ramond charges that can be reinterpreted as open strings ending on a $p$-brane with Dirichlet 
boundary conditions, the so-called D-brane [9] . D-branes provide a powerful tool to study 
nonperturbative properties of superstring theories and also admit a second class of BPS states that appear 
after compactification of type II and type I strings on a Calabi Yau space. Furthermore, 
there are world volume vector fields in  D-branes actions.

To be more precise, the low energy physics, see reference [13] for an extensive review ,  of $N$ $coincident$ Dirichlet $p$-branes in $10D$ flat space is described in the static gauge ( identifying the world-volume coordinates of the $D$-brane with $p+1$ of the ten dimensional flat space coordinates) by the dimensional reduction to $p+1$ dimensions of a $U(N)~N=1$ SYM in 
$D=10$. $1/2$ of supersymmetries are broken and consequently there are $8$ on-shell bosonic and $8$ on-shell fermionic degrees of freedom. Classically it is a well defined theory but quantum mechanically is anomalous. In particular, the low energy dynamics of $N$ $D0$-branes  
in the gauge where the gauge field $A_0=0$ is given by the dimensionally reduced SYM theory from $10D$ to $0+1$ dimensions. The Lagrangian is  :

$${1\over 2g\sqrt {\alpha '} } Tr [ (\partial_t X^I)(\partial _t X_I) + {1\over (2\pi \alpha')^2} 
\sum [X^I, X^J] ^2 + {1\over 2\pi \alpha'} i\theta^T (\partial_t \theta) -{1\over (2\pi \alpha ')^2} 
\theta^T \Gamma_I [X^I, \theta] ]. \eqno (2-13)$$
Each of the nine adjoint scalar ( from the point of view of the world volume)  matrices $X^I$ is a hermitean $N\times N$ matrix, where $N$ is the number of $0$-branes. The $\theta$ are $16$-component spinors which transform under the $SO(9)$ Clifford algebra given by the $16\times 16$ matrices $\Gamma_I$. 

The $N\rightarrow \infty$ limit of  (2-13) ( see [14] for an update and [15] for a textbook )  can be obtained by replacing the adjoint scalar matrices $X^I$ by $c$-number functions $X^I(t,\sigma^1,\sigma^2)$ and matrix-commutators by Poisson brackets w.r.t two internal coordinates $\sigma^1,\sigma^2$ and the group trace 
by an integral w.r.t the internal coordiantes $\sigma^1,\sigma^2$. In this fashion one obtains a gauge theory of symplectic diffs of a two-dim surface, a membrane. 
The time integral in conjuction with the integration  w.r.t the $\sigma^1,\sigma^2$ variables yields in effect an action similar to the light-cone gauge action ( excluding the zero modes) for a supermembrane of spherical topology moving in a flat target spacetime background. Concentrating solely on the bosonic sector, the $N\rightarrow \infty$ limit of 
(2-13) yields :

$$ lim_{N\rightarrow \infty}  S_{Bosonic} =\int dt \int d^2\sigma ~  {1\over 2g\sqrt {\alpha '} } [ (\partial_t X^I)(\partial _t X_I) + {1\over (2\pi \alpha')^2} 
\sum \{X^I, X^J\} ^2] . \eqno (2-14). $$
We see  that (2-14) is similar to the light-cone gauge spherical membrane action in flat $D=11$. The latter action ( excluding the zero modes) is equivalent to a $10D$ YM action for the $SU(\infty)$ group dimensionally reduced to one temporal dimension. Whereas, (2-14) represents the low energy dynamics of $N=\infty$ coincident $DO$-branes  and was obtained from a dimensional reduction of a $10D$ YM theory to one temporal dimension, after  fixing the gauge $A_0=0$. The $9$ scalars $X^I (t,\sigma^1,\sigma^2)$ are transverse to the world volume of 
the $N=\infty$ coincident $D0$-branes  in a $10D$ flat space and are obtained from  the decomposition of the 
$10D$ YM field $A_M$ into a $p+1$-dim gauge field $A_\mu$ and $10-(p+1)=9-p$ transverse scalars. In the special case that $p=0$ the gauge field $A_\mu=A_0$ and the $9-0=9$ scalars are the $X^I$ transverse coordinates of the $D0$-branes.     
To sum up, the low energy dynamics of an infinite number of coincident $D0$-branes in flat $10D$ space resemble those of a lightcone spherical membrane in flat $11D$ space.  The membrane ground state can be ``seen''  as a condensation of an infinity of $D0$-branes. Since the action given eq-(2-14) has a similar  form to the generalized Skyrmion action of eqs-(2-12a,2-12b) we suggest  that $D0$-branes ought to play an important part in the spectrum of the Skyrmion action in the large $N\rightarrow \infty$ limit.

Solutions to actions of the form (2-14) in $8D$ have been studied by Ivanova-Popov  [16] and Fairlie, Ueno [17] ( octonionic Nahm equations)   
and in particular can be reducced to the ordinary Nahm equations in $4D$ which admit 
BPS monopole solutions. Massive BPS states appear  in theories with $extended $ supersymmetry but even in the case of $N=1$ supersymmetry there is an analog of BPS sates, namely the massless 
states [18]. A future proyect is to see what connection there may be, if any,  among 
these massless sates, the analogs of BPS states, with the infinite number of coincident $D0$-branes and the spectrum of the generalized Skyrmion action.    
Hints that a relation exists are based on the fact that :
(i) we have a world-volume vector field $b_\mu$ in our Weyl invariant spinning membrane action 
in contrast with the 
$p+1$-dim $U(N)$ gauge field $A_\mu$ living on the world-volume of a $Dp$-brane.   
(ii) Weyl conformal invariance (iii) $Q$  supersymmetry only ( 1/2 supersymmetry, as it occurs in BPS states) and no 
$S$ supersymmetry.     
(iv) The Skyrmion action (2-12a,2-12b) subsumes the action for the large $N\rightarrow \infty$ coincident $D0$-branes given by eq-(2-14).  
  
\bigskip

\centerline{\bf { III}. The $Q$-spinning Weyl Invariant Membrane}
\bigskip
\centerline{\bf Supersymmetrization of the Kinetic Terms}
\bigskip

    In this section we will present an action for the $3D$ Kinetic matter
superconformal multiplet where supersymmetry is linearly realized and without $R$ terms.
Also we will supersymmetrize the quartic-derivative terms of (2-7). This is
attained by using directly an explicit  superconformally invariant action for
the kinetic terms. The quartic terms do not admit a superconformally invariant
extension unless one includes a suitable coupling as we shall see shortly. The key issue lies in the fact that if we
wish to satisfy the three requirements: 

1). A spinning membrane action which is polynomial in the fields.

2).  Absence of $R$ terms.

3). Linearly realized supersymmetry in the space of fields after the
elimination of the auxiliary fields, before and after one sets the Fermi fields
to zero.
One must relinquish $S$ supersymmetry altogether and concentrate solely on the
$Q$ supersymmetry associated with the superconformal algebra in D=3. We shall
begin with some definitions of simple-conformal SG in D=3 [4]:

The scalar and kinetic multiplet of simple conformal SG in $3D$  are respectively:

$$\Sigma_c =(A,\chi,F).~~T_c(\Sigma_c)=(F,\gamma^\mu D_\mu^c \chi, \triangle  A)\eqno(3-1){ }$$

We have the following quantities:

$$ D^c_\mu A =\partial_\mu A -{1\over 2} {\bar \psi}_\mu\chi -\lambda b_\mu A.                                              \eqno(3-2){ }$$

$$  D^c_\mu \chi =(D_\mu -(\lambda +{1\over 2})b_\mu )\chi -{1\over 2}\gamma^\mu D_\mu^c A\psi_\mu 
-{1\over 2}F\psi_\mu -\lambda A\phi_\mu.
\eqno(3-3){ }$$

$$\triangle  A =D^c_a D^{ca} A=e^{-1} \partial_\nu (eg^{\mu\nu}D^c_\mu A)+
{1\over 2}{\bar \phi}_\mu \gamma^\mu \chi -3\lambda b^\mu D^c_\mu A +$$
$$2\lambda A f^a_\mu e^\mu_a -{1\over 2}{\bar \psi}^\mu D^c_\mu \chi 
-{1\over 2}{\bar \psi}^\mu \gamma^\nu \psi_\nu D^c_\mu A. \eqno(3-4){ }$$

$$\omega^{mn}_\mu =-\omega^{mn}_\mu (e)-\kappa^{mn}_\mu (\psi)+e^n_\mu b^m -e^m_\mu b^n.
\eqno(3-5-a){ }$$

$$\phi_\mu ={1\over 4} \gamma^\lambda \gamma^\sigma S_{\sigma \lambda} =
{1\over 4} \sigma ^{\lambda\sigma}  \gamma_\mu S_{\sigma \lambda}. 
\eqno(3-5-b){ }$$

$$\kappa^{mn}_\mu ={1\over 4} ({\bar \psi}_\mu \gamma^m \psi^n -{\bar \psi}_\mu \gamma^n \psi^m 
+ {\bar \psi}^m \gamma_\mu \psi^n). 
\eqno(3-5-c){ }$$

$$S_{\mu\nu} =(D_\nu +{1\over 2} b_\nu )\psi_\mu - \mu \leftrightarrow \nu. 
\eqno(3-5-d){ }$$

$$e^{a\mu} f_{a\mu} =-{1\over 8} R(e,\omega) -{1\over 4}{\bar \psi} _\mu \sigma^{\mu\nu}\phi_\nu. 
\eqno(3-5-e){ }$$

         The transformation laws under Weyl scalings, Q and S supersymmetry are
respectively:

$$ \delta e^m_\mu =\lambda e^m_\mu.~~~\delta A={1\over 2}\lambda A.~~~\delta \chi=\lambda \chi.~~~
\delta F={3\over 2}\lambda F
\eqno(3-6)$$

$$\delta^c_Q A ={\bar \epsilon} \chi.~~~\delta^c_Q \chi =F\epsilon+\gamma^\mu D^c_\mu A\epsilon. ~~~
\delta^c_Q F={\bar \epsilon}\gamma^\mu D^c_\mu \chi.
\eqno(3-7)$$

$$\delta^c_Q e^m_\mu ={\bar \epsilon}\gamma^m \psi_\mu.~~~
\delta^c_Q \psi_\mu  =2( D_\mu +{1\over 2}b_\mu )\epsilon. ~~~
\delta^c_Q b_\mu = \phi_\mu.
\eqno(3-8)$$
The S-supersymmetry transormations are :

$$\delta^c_S e^m_\mu =0.~~~ \delta^c_S \psi_\mu =-\gamma_\mu \epsilon_s.~~~ 
\delta^c_S b_\mu =-{1\over 2}\psi_\mu \epsilon_s.  \eqno (3-9a) $$       

$$\delta^c_S \omega^{mn}_\mu =- {\bar \epsilon}_s \sigma^{mn} \psi_\mu.$$

$$\delta^c_S A =0. ~~\delta^c_S \chi =\lambda A \epsilon_s.~~~
\delta^c_S F = ({1\over 2}-\lambda) {\bar \chi} \epsilon_s. \eqno (3-9b)$$

The kinetic multiplet transforms propely under
Q-transformations for any value of the conformal weight but $not$  under S supersymmetry 
transformations unless one assigns the canonical weight  
$\lambda ={1\over 2}$ to the $\Sigma_C$ supermultiplet ( to its first member, $A$,  so that 
$F$ has a weight of $1+{1\over 2}$ and 
the associated  kinetic  
multiplet also has a Weyl weight equal to ${3\over 2}$ ).  
A superconformally invariant action for the kinetic terms is  [4] :

$${L=e[{\hat F}+{1\over2}\bar{\psi_\mu}\gamma^\mu {\hat \chi}
+{1\over2}{\hat A}\bar{\psi_\mu}\sigma^{\mu\nu}\psi_\nu].}\eqno(3-10)$$  
where one inserts the multiplet $\Sigma_C\otimes T_C(\Sigma_C)=( {\hat A}, {\hat \chi} ,{\hat F})$ into (3-10). 
One  must make sure  to have $\lambda  = {1\over 2}$ for $\Sigma_C$ so that the ${\hat F} $ component 
appearing in (3-10) has dimension three otherwise we would not
even have Q-invariance in the action   despite the fact that the kinetic multiplet transforms
properly under Q-transformations irrespectively of the value of $\lambda $.

On physical grounds we see that the notion of canonical dimension is
intrinsically tied up with the conformal invariant aspect of the kinetic terms
in the action. We have a conformally invariant kinetic term if, and only if, the
fields have the right (canonical) dimensions to yield terms of dimension three in
the Lagrangian. The auxiliary field $F$ appearing in the action has the desired quadratic $F^2$ pieces  that allows it to be 
eliminated algebraically without introducing constraints among the $A,\chi$ fields. Upon the elimination of $F$ yields $F=F(A,\chi)$. It is true that if one were to fix the Weyl invariance in the last 
relation by setting $F=constant$ this will reintroduce constraints; 
however these are $not$ entirely due to the algebraic elimination of the auxiliary field, $F$, but to 
the fact that a gauge condition has been selected. A gauge choice naturally constrains fields or some 
of its components. We shall go back to this point later.

We might ask ourselves how did Lindstrom \& Rocek manage to
construct a Weyl invariant spinning membrane when their fields had a
non-canonical dimension? The answer to this question lies on the nonpolynomial
character of their action. Formally one has an infinite series expansion where the whole sum of 
explicitly $Q$ and $S$ supersymmetry breaking terms  is effectively invariant under the $''Q+S''$ 
sum rule.  
An example of a multiplet that transforms properly under the $''Q+S''$ sum rule but $not$ separately 
under $Q$ nor $S$ supersymmetry is the following Poncare kinetic multiplet :

$$T_p(\Sigma_p)=(F; D^c\chi (\lambda ={1\over2}); \triangle A -{3\over 4}FS
).\eqno (3-11)$$
This multiplet is almost identical as the kinetic superconformal multiplet (3-1) 
except that the last component is different due to the presence of the $-{3\over 4} FS$ term. 

The task now is to see how do we write a suitable action for the kinetic
terms without $R$ terms ( which appear in the definition of the D'Alambertian)
for values of $\lambda$ different than zero. The suitable action is obtained
as follows:

Take the   combination $\Sigma^i_C \otimes T_C(\Sigma^j_C ) + T_C(\Sigma^i_C
)\otimes \Sigma^j_C -
   T_C( \Sigma^i_C \otimes \Sigma^j_C)$ which happens to be the correct one to
dispense of the $R$ terms. The explicit components of the latter multiplet are :

$$A_{ij}= \bar{\chi_i} \chi_j.\eqno(3-12a)$$                                  

$$\chi_{ij} =F_i\chi_j +F_j\chi_i +A_i\gamma^\mu D^c_\mu (\lambda ={1\over 2})\chi_j  
+A_j\gamma^\mu D^c_\mu (\lambda ={1\over 2})\chi_i-$$
$$\gamma^\mu D^c_\mu (\lambda =1)[A_i\chi_j + A_j\chi_i].        
\eqno(3-12b)$$            

$$F_{ij}= A_i \triangle (\lambda ={1\over 2})  A_j +A_j \triangle (\lambda ={1\over 2})A_i +2F_iF_j 
-{\bar \chi}_i \gamma^\mu D^c_\mu ( \lambda ={1\over 2})\chi_j -$$      
$$  {\bar \chi}_j \gamma^\mu D^c_\mu ( \lambda ={1\over 2})\chi_i - \triangle ( \lambda =1)[A_iA_j]. \eqno (3-12c)$$       

The $F_{ij}$ terms contain the standard kinetic terms : 

$$-2g^{\mu\nu}\partial_\mu A_i\partial_\nu A_j -2\bar{\chi_i}\gamma^\mu D_\mu\chi_j
+2F_iF_j+......\eqno (3-12d)$$ 
and $no$ curvature terms by construction. 
Unfortunately matters are not that simple! It is true that the
components of the latter mutiplet transform properly under Q transformations
since each single one of the components in the definition of
eqs-(3-12) does. However, this not the case for S-supersymmetry since
the component, $T(\Sigma\otimes\Sigma)$, does  $not$ transform properly under S-supersymmetry
because the weight of $\Sigma\otimes\Sigma$  is equal to $1$ instead of
${1\over2}$. Therefore, eliminating the $R$ terms is $not$ compatible with
S-supersymmetry. We are force, then, to relinquish S-supersymmetry and
implement Q-supersymmetry only. 
   
     Our action is invariant under Q-supersymmetry and is obtained by plugging in directly
$A,\chi$ and $F$ obtained in eqs-(3-12)  into  eq-(3-10) and contracting the spacetime indices with
$\eta_{ij}$ . It has a similar form as (2-10) but it does $not$ contain the term
linear in $S$, $S\bar{\chi}\chi$, exclusively , which was the one which furnished
the constraint between our physical fields in (2-10) after  elimination of
$S$ . 
Moreover, we don't have $R$ terms, and Q-supersymmetry is linearly realized after the
elimination of $F^i$. Notice the explicit presence of the $b_\mu$ terms in eq-(3-12c) resulting from the definition of 
the D'Alambert operator in eq-(3-4). This is tied up 
with the fact that there are no scalar curvature terms after one computes (3-12c). 
In $4D$ we learnt that the Weyl field 
$A_\mu$ decouples in the expression 
$$(D^{Weyl}_\mu D^{\mu} +\lambda R^{Weyl})\phi =0 . \eqno (3-13)$$
if one choosses the coupling $\lambda ={1\over 6}$. This is due to an exact cancellation between the 
$A_\mu$ field appearing in the Weyl scalar curvature and the D'Alambertian. It is no surprise that if 
the curvature is eliminated in (3-12-c) one will have $b_\mu$ terms remaining. However, Weyl covariance is maintained !
Eq-(3-4) that defines the three-dim version of the D'Alambert operator is Weyl covariant by $construction$. Notice the 
$explicit$ presence of the $-3\lambda b^\mu D^c_\mu A$ term in the r.h.s of (3-4). Under Weyl transformations the 
inhomogeneous pieces will cancel those stemming from the first term in the r.h.s of (3-4) 
$e^{-1}\partial_\nu (eg^{\mu\nu}D^c_\mu A)$. To have an explicit $b_\mu$ dependence does not imply that Weyl covariance
is destroyed. The tensor calculus does not break Weyl covariance. Eqs-(3-12) are defined explicitly in terms of the tensor products.    
 
\centerline{\bf Supersymmetrization of the Quartic Derivative Terms}
\bigskip

 We  proceed now to supersymmetrize the $L_4$ terms. Unless one introduces 
a coupling multiplying the quartic derivative bosonic terms, one cannot
obtain a superconformally invariant action (not even Q-invariant)  now because
these terms do not have the $net$ conformal weight of $\lambda =2$ as  the kinetic
terms had. (We refer to the net weight of the first component of a multiplet so
that  $F$  has dimension three). For this reason we have to introduce the
following coupling function, a multiplet, that has no dynamical degrees of
freedom but which serves the purpose of rendering the quartic-derivative terms
with an overall dimension three to ensure that our action is in fact
dimensionless. We refrained from doing this sort of "trick" in the case of the
kinetic terms because such terms are devoid of a dimensional coupling constant.
The Dolan Tchrakian  action contains an arbitrary constant in front of the
quartic pieces and it is only the ratio between this constant and the
dimensionless constant in front of $L_2$ which is relevant. This constant must
have dimensions of $(length)^{3}$ since we have an extra piece of dimension
three stemming from the term ,$(\partial_\mu A)^2 $.

     Let us, then , introduce the coupling-function supermultiplet, 

$$\Sigma_0=(A_0;\chi_0;F_0).$$
whose Weyl weight is equal to $-3$  so that the tensor product of $\Sigma_0$ with
the following multiplet, to be defined below, has a conformal weight ,
$\lambda=2$ as it is required in order to have Q-invariant actions.

Lets introduce the following multiplet ( with an overall $\lambda =5$ so the $F$ terms have dimension six). 
$$K^{ijkl}_{\mu\nu\rho\tau}\eta_{ij}\eta_{kl}=K(\Sigma^i_\mu;\Sigma^j_\nu)
\otimes{T[K(\Sigma^k_\rho;\Sigma^l_\tau)}] +({ij\leftrightarrow kl})~
and~({\mu\nu\leftrightarrow\rho\tau})-$$

$$T[K(\Sigma^i_\mu;\Sigma^j_\nu)\otimes
{K(\Sigma^k_\rho;\Sigma^l_\tau)}]\eta_{ij}\eta_{kl}. \eqno (3-14a)$$
                  
$$K(\Sigma ,\Sigma)=\Sigma^i_C \otimes T_C(\Sigma^j_C ) + T_C(\Sigma^i_C
)\otimes \Sigma^j_C -
   T_C( \Sigma^i_C \otimes \Sigma^j_C). \eqno (3-14b)$$
which is the adequate one to retrieve (2-7) at the bosonic level and also the
one which ensures that the $R$ terms do cancel from the final
answer. This is indeed the case as it was shown in eqs-(3-12). A 
calculation yields the components of the supersymmetric-quartic-derivative
terms: 

$$A_{ijkl}= \bar{\chi_{ij}  } \chi_{kl}.\eqno(3-15a)$$                                  

$$\chi_{ijkl} =F_{ij}\chi_{kl} +F_{kl}\chi_{ij} +A_{ij}\gamma^\mu D^c_\mu (\lambda ={2})\chi_{kl}  
+A_{kl}\gamma^\mu D^c_\mu (\lambda ={2})\chi_{ij}-$$
$$\gamma^\mu D^c_\mu (\lambda =4)[A_{ij}\chi_{kl} + A_{kl}\chi_{ij}].        
\eqno(3-15b)$$            

$$F_{ijkl}= A_{ij} \triangle ( \lambda = 2)  A_{kl}  +A_{kl}  \triangle (\lambda = 2 )  A_{ij}   +2F_{ij} F_{kl} 
-{\bar \chi}_{ij}  \gamma^\mu D^c_\mu ( \lambda ={2})\chi_{kl}  -$$      
$$  {\bar \chi}_{kl}  \gamma^\mu D^c_\mu ( \lambda = 2)\chi_{ij}  - \triangle ( \lambda = 4)[A_{ij}A_{kl}]. \eqno (3-15c)$$       
  
Where we have used the abbreviations $A_{ij},\chi^{ij}$ and
$F^{ij}$ already given in eqs-(3-12) and where the derivatives acting on composite fields must appear with the 
right Weyl weight. 
Notice the similarity ( as one should have) between eqs-(3-12) above and eqs-(3-15), both in
form and in the values of the coefficients. This it due to the tensor calculus nature of the 
supermultiplets. This is a sign of consistency  and should serve as a check. We
need to take the tensor product of the latter multiplet given above in eq-(3-15) and the
coupling-function multiplet:
$$\Sigma_0\otimes(A^{ijkl};\chi^{ijkl};F^{ijkl})=(A_0A^{ijkl};A_0\chi^{ijkl}+\chi_0A^{ijkl};
A_0F^{ijkl}+F_0A^{ijkl} -{\bar \chi_0}\chi^{ijkl}). \eqno (3-16)$$
The complete Q-supersymmetric extension of $L_4$ requires adding terms
which result as permutations of ${ijkl\leftrightarrow ilkj\leftrightarrow
kjil\leftrightarrow klij}$ keeping $\eta_{ij}\eta_{kl}$ fixed and inserting (3-16) into (3-10).
One has : (i) Linearly realized supersymmetry( Q-supersymmetry).
(ii) Absence of $R$ terms.
(iii) A polynomial Lagrangian in the fields.

Eliminating the auxiliary fields, ${\partial L/{\partial F^i}}=0$,
and setting the Fermi fields to zero we must recover the Weyl-covariantized Dolan-Tchrakian 
action (WCDT). Furthermore, the order in which we perform this should yield identical
results: set the Fermi fields to zero and eliminate the auxiliary fields and vice versa.

    It is fairly clear that we have enforced Q-supersymmetry. The fields which
comprised the "coupling" function do not explictly break conformal invariance
and are not to be varied ( one does not vary couplings ordinarily) otherwise we would constrain the supersymmetrization of the quartic terms in 
the action to zero as discussed in the introduction. The members of the 
coupling function supermultiplet are not Lagrange multipliers enforcing any constraints and for this reason these couplings must not be varied. A question arises : If one does not vary the couplings in the action how then can one determine their values ? 

This was thoroughly explained earlier in the introduction. Fistly, one eliminates algebraically the members of the gravitational supermultiplet via their variation in the action ( the gravitational supermultiplet is not dynamical ) and, afterwards, one imposes the embedding condition : $D_\mu A_0=0$. Upon supersymmetry transformations, the remaining members of the coupling supermultiplet are eliminated as well.  In this way one will be able to express the couplings and the members of the gravitational supermultiplet in terms of the membrane matter fields and their superpartners. Further details of how this is attained are discussed in the Appendix. 
 
Eliminating the $F^i$ field yields the value for $F^j$=0, after
we set the Fermi fields to zero. Conversely, setting the fermions to zero and eliminating the 
$F^i$ fields yields zero as the viable solution for the $F^i$ fields since we don't wish to 
generate constraints
among our physical fields. It is precisely when we set the Fermions to zero that the
$F_0A^{ijkl}$ term vanishes and the $\bar {\chi_0}  \chi^{ijkl}$ terms as well. We
are left only with the $A_0F^{ijkl}$ piece belonging to the bosonic quartic terms in the WCDT action as 
intended. 
Finally, fixing the Weyl gauge invariance by setting $A_0=g$ renders the WCDT action in the original DT form once the 
embedding condition $D^{Weyl}_\mu A_0 =0 $ is implemented ( once the fermions are set to zero the 
superconformal derivative becomes the ordinary Weyl derivative). $A_0=g \Rightarrow b_\mu \sim \partial_\mu lnA_0=0$. 

Had one fixed the Weyl invariance firstly in the $Q$ invariant action one would encounter constraints 
among the physical fields after eliminating the $F^i$ auxiliary fields and members of the 
gravitational  supermultiplet, if, and only if,  the embedding condition is maintained. 
However, these constraints are  due to the gauge fixing and not as a result that the supersymmetry is realized nonlinearly. We have discussed earlier how upon eliminating the $F$ field in the 
kinetic superconformally invariant action 
(3-10), $F=F(matter)$ and fixing the Weyl gauge invariance by setting  $F=constant$ 
afterwards, will naturally introduce constraints. These are constraints as a result of the gauge fixing.

Therefore, one $must$ have conformal 
invariance in the $Q$ spinning membrane otherwise constraints will re-appear among the physical 
fields. This was the whole point of embedding the DT action into a superconformally invariant one 
and behind the Lindstrom-Rocek construction. There has been a trade off between $b_\mu \leftrightarrow 
{1\over 4} \gamma_\mu S$. It was the elimination of $S$ in Poincare Supergravity which generated constraints and yielded a nonlinear realization of supersymmetry.

To conclude we have Q-supersymmetrized the Dolan-Tchrakian action. The
kinetic terms and quartic terms are Q-invariant by construction. The latter
ones were Q-invariant with the aid of an extra multiplet, the
"coupling" function multiplet whose weight is precisely equal to $-3$ to ensure
that our action is dimensionless and scale invariant. After 
eliminating the $F^i$ auxiliary fields, having set the
Fermi fields to zero, 
and fixing the dilational invariance 
we retrieve the Dolan-Tchrakian Lagrangian for the membrane. The main
point of this paper is to show that one can have a Q-spinning membrane 
solely if we wish to satisfy all of the requirements listed in the introduction.
"Q+S" invariance can only be implemented in non-polynomial actions as Rocek and
Lindstrom showed [5].

Since the only obstruction to fixing the gauge $b_\mu\sim \partial_\mu lnA_0 =0$ globally is topological
it is warranted to study the topological behaviour of these 3-dim gauge fields
and see what connections these may have with Witten's Topological QFT, 
Chern-Simmons 3-dim Gravity and with other non-perturbative phenomena in three
dimensions [7].

\centerline {\bf Acknowledgements}
\medskip
We are indebted to the Center for Theoretical Studies of Physical Systems for support and to Luis Boya for
discussions. Also to Yuval Ne'eman for having suggested this problem and 
George Sudarshan for his help.

\medskip
\centerline { \bf {APPENDIX}}														
Now we turn to the discussion concerning the presence of the $b_\mu $ 
which is crucial since now we do not have at our disposal the possibility of
fixing the $K$-invariance to set $b_\mu$  =0. We have decided to include this
discussion in the following Appendix because the whole essence of this last
section has been based on Q-invariance.
The discussion in section III cannot be complete unless we study in
detail the behaviour of our final action due to the presence of the $b_\mu$
terms. We have two cases to consider:

1-. The case where $b_\mu$ decouples from the action and from the
Q-supersymmetry transformation laws of the action. An example of this is 
the general form of an action given by eq-(3-10) for the particular case that one chooses the multiplet
$ \Sigma_C\otimes { T_C(\Sigma_C)} $ with a net weight equal to $\lambda=2$. One can
see , explicitly, by inspection 
that the $b_\mu$ terms decouple. When there is no explicit $b_\mu$
dependence in the action (3-10) one has  implemented $K$ invariance.
Furthermore, under Q-supersymmetry , eq-(3-10)
contains the following terms:

$\delta F$ yields a term $ \bar{\epsilon}\gamma^\mu (D_\mu -(2+{1\over 2})
b_\mu)\chi$ whose $b_\mu$ term  is

$$-(2+{1\over2})\bar{\epsilon}\gamma^\mu b_\mu \chi.\eqno (A-1)$$
A factor of $-b_\mu$ cancels against the $b_\mu$ terms contained in the
$\omega(e;\psi;b_\mu)$ leaving us with a net factor of $-{3\over2}b_\mu$. Whereas the
second term of (3-10) yields , upon variation of the gravitino using (3-8), the
following term:
${1\over2}\bar{\epsilon}\gamma^\mu b_\mu\chi $
plus another factor of $+b_\mu$ stemming from the spin-connection leaving a
net factor of ${3\over2}b_\mu $. The latter factor will cancel the previous $-{3\over2}b_\mu$ factor .
It is clear that $b_\mu$ does also decouples from the
transformation laws.

2-. The case when $b_\mu$ does not explicitly decouple from the action but it does from
the $Q$ transformation laws to ensure Q-invariance. Since we can no longer choose
the gauge $b_\mu=0$ these $b_\mu$ terms must cancel out in the transformation laws because Q-invariance was
not broken explicitly . 
This is our case. We bring to the attention of the reader that
$ \Lambda=T(\Sigma\otimes {\Sigma})$ is not K-inert. The simplest way to see
this is by looking at the superconformal algebra in three dimensions.
$$ [\Lambda,[Q,K]] +[Q,[K,\Lambda]] +[K,[\Lambda,Q]] =0.\eqno (A-2) $$
Since
$ [\Lambda,Q] =0, [Q,K] \sim {S}$ and $[\Lambda, S]\not= 0$ we have
$ [K,\Lambda]\not=0.$
Therefore $K$ symmetry is broken explicitly and we have $b_\mu$ terms in our final expressions for 
the action.

A rough illustration of the type of terms to be studied after eliminating the auxiliary $F^i$ fields and 
setting the fermions to zero 
is the following. 
The Weyl covariantized DT action 
has the form ( we are not including the target spacetime indices nor the antisymmetrization 
of the $3D$ indices as well):
$$-(\partial_\mu A -\lambda b_\mu A)^2 +A_0(\partial_\mu A -\lambda b_\mu
A)^4.\eqno (A-3) $$
where $A_0 <0$ since from (2-7) we know that the relative sign between the quadratic and quartic terms 
is the same. Eliminating the $b_\mu$ after a variation w.r.t the $b_\mu$ 
yields:
$$  (D_\mu^{Weyl} A) +2(-A_0)(D_\mu^{Weyl} A )^3 =0. \eqno (A-4) $$
where we have factored  out the term $-\lambda A $ (which should not be
constrained to zero). therefore
one gets two possible solutions:
$$ 1-2A_0(D_\mu ^{Weyl} A )^2 =0 \Rightarrow (D_\mu ^{Weyl} A )^2 ={1\over 2 A_0}<0. \eqno (A-5) $$
this is consistent with the timelike condition of the vector $(D_\mu ^{Weyl} A$). 
The other condition is 
$(D_\mu ^{Weyl} A )=0$ which is unacceptable because it  
constrains the action to zero which is not very physical  whereas the former condition is fine. 
Implementing the    
embedding condition $(D_\mu ^{Weyl} A_0 )=0\Rightarrow b_\mu \sim \partial_\mu ln A_0$ in (A-5) yields
 the 
desired relationship among $A_0,A,b_\mu$ :

$$(D_\mu ^{Weyl} A )^2 = [\partial_\mu A - \lambda (-{1\over 3} \partial_\mu ln A_0) A]^2 =
{1\over 2 A_0}. \eqno (A-6)$$
The last equation yields the relation between $A_0$ and  $A$ and the relation $b_\mu =-{1\over 3} (\partial_\mu ln A_0)$
determines $b_\mu$ in terms of $A$. In this fashion one has found the relationship among all the fields in terms of the physical membrane coordinates. It is true that (A-6) is not an algebraic relation 
between $A_0,A$, however this does not spoil the linear realization of supersymmetry among the membrane's fields. Once more, if one were to fix the Weyl invariance by setting $A_0 =g$ constraints will 
reappear among the latter fields but due to the gauge choice condition and not entirely as a 
result of the 
elimination of the auxiliary fields.    

The construction in section III presupposes the fact that one can find a gauge where
(simultaneously) the scalar field 
$A_0$  can be gauged to a constant and the
$ b_\mu$ field to zero. The equations of motion of the $A$ fields stemming from the quartic derivative 
terms of the WCDT action are of the form :

$$D^{Weyl}_\mu [{\delta S_4 \over \delta (D^{Weyl}_\mu A)}] 
\sim (D_\mu ^{Weyl} A_0 )(D_\mu ^{Weyl} A )^3
+A_0 D^{Weyl}_\mu [(D_\mu ^{Weyl} A )^3]. \eqno (A-7)$$

The above expression is Weyl covariant and we have assumed that there are no boundary terms in our action and that the fields vanish fast enough at infinity....As it is usual in these variational problems we have integrated by parts and generalized Stokes law to the Weyl space. 
Now we can derive the embedding condition from first principles. If the equations of motion of the WCDT action are indeed the Weyl covariant extension of the equations of motion of the DT action then the condition $(D_\mu ^{Weyl} A_0 )=0$ follows immediately since there is no analog of that term in the DT equations of motion. A similar argument follows in the superconformal case by replacing the Weyl covariant 
derivative by the superconformal one :  $(D_\mu ^c A_0 )=0$.

To finalize this Appendix we point out that the only obstruction in setting
$b_\mu$ to zero must be topological in origin. We saw in section II
that it was the elimination of S which originated the constraint
$\bar{\chi}\chi$ =0. Such S term had the same form as an effective mass resulting from a 
fermion-condensate.
Whereas here, upon the "trade-off" $ b_\mu \leftrightarrow {{1\over4}\gamma_\mu S}
$ we may  encounter topological obstructions in setting $b_\mu$ =0 gobally
and, henceforth, in Q-supersymmetrizing the Dolan-Tchrakian action; i.e.  to
obtain the exact bosonic limit from the Q-supersymmetric action. 
\medskip

\centerline {\bf References}
\smallskip
1-.P.S Howe and  Tucker, J. Math. Physics {\bf 19} (4) (1978) 869.

J.Math. Physics {\bf 19}  (5) (1978) 981. J. Physics {\bf A 10} (9) (1977) L 155.

2-.M.Duff,  "Supermembranes, The first 15 Weeks ". CERN-TH- 4797 (1987).

Class. Quan. Grav {\bf 6} (1989) 1577. 

3-.E.Bergshoeff, E.Sezgin and P.K.Townsend,  Phys.  Lett.  {\bf B 209} (1988) 451.

4-.T.Uematsu,  Z.Physics {\bf C 29}  (1985) 143-146 and {\bf C32}  (1986) 33-42.

5-. U.Lindstrom,  M. Rocek,  Phys. Letters {\bf B 218}  (1988) 207.

6-.B.P.Dolan,  D.H.Tchrakian,  Physics Letters {\bf B 198}  (1987) 447.

7-S.Deser, R.Jackiw and  S. Templeton: Ann. Physics {\bf 140} (1982) 372.

8-C. Castro, Journal of Group Theory in Physics {\bf 1} (2) (1993) 215.   

J. Chaos, Solitons and Fractals, {\bf 7} (5) (1996) 711.

9-J. Polchinski , TASI lectures on D-branes, hep-th/9611050. 

J. Polchinski , S. Chaudhuri and C. Johnson : `` Notes on D-branes, hep-th/9602052. 

10-E. Witten, Nucl. Phys. {\bf B 443}  (1995) 85.

C. Hull and P. Townsend, Nucl. Phys. {\bf B 438} (1995) 109.

C. Vafa, `` Evidence for F theory `` , hep-th/9602022. 

J.Schwarz : `` The Power of M theory'' , hep-th/9510086

R. Dijkgraaf : `` Les Houches Lectures on Fields , Strings and Duality ``,  

hep-th/9703136

11- T.H.R.  Skyrme , Proc. Roy. Soc. {\bf A 260} (1961) 127. 

C. Houghton, N. Manton, P. Sutcliffe, `` Rational Maps, Monopoles and Skyrmions ``.

hep-th/9705151. 

N. Manton, Comm. Math. Phys {\bf 111} (1987) 469. 

M. Duff, S. Deser and C. Isham, Nucl. Phys. {\bf B 114}  (1976) 29.

12- T. Banks, W. Fischler, S. Shenker and L. Susskind, 

`` M theory as a Matrix model, a Conjecture `` hep-th/9610043. 

Phys. Rev {\bf D 55} (1997) 5112.

13- W. Taylor : `` Lectures on $D$- branes, Gauge Theory and M(atrices)'' 

hep-th/9801182. 

14-   C. Zachos, D. Fairlie and T. Curtright : `` Matrix Membranes and Integrability `` 

hep-th/ 9709042.

15-Y. Ne'eman , E. Eizenberg : `` Membranes and Other Extendons ( p-branes) `` 

World Scientific Lecture Notes in Physics {\bf vol 39} . (1995).

16- Ivanova, Popov : Jour. math. Phys. {\bf 34} (2) (1993) 674. 

Theor. Math. Phys. {\bf 94} (2) (1993) 316. 

17. D. Fairlie, T. Ueno : `` Higher Dimensional Generalizations of the Euler 

Top Equation `` hep-th/9710079.

T. Ueno : `` General Solution to the $7D$ Top Equation `` hep-th/9801079.

18. E. Kiritsis : `` Introduction to Non-Perturbative String Theory `` 

hep-th/9708130.

\bye